\documentclass[jctcce,manuscript=article]{achemso}

\usepackage{graphicx,bm,xcolor,microtype,multirow,amscd,amsmath,amssymb,amsfonts,physics,longtable,wrapfig,txfonts,soul}
\usepackage{lineno,appendix}

\usepackage[utf8]{inputenc}
\usepackage[T1]{fontenc}
\usepackage{txfonts}
\usepackage{dsfont}
\usepackage{framed}
\usepackage[normalem]{ulem}

\usepackage[
	colorlinks=true,
    citecolor=blue,
    breaklinks=true
	]{hyperref}
\usepackage{scalerel}

\newcommand{\sys}{S}
\newcommand{\env}{E}

\newcommand{\Hp}{}  

\def\myhrulefill{\hspace*{\parindent}\leavevmode\leaders\hrule\hfill\kern 0pt}

\newenvironment{myleftbar}{%
  \MakeFramed {\advance\hsize-\width \FrameRestore}}%
 {\endMakeFramed}

\makeatletter
\def\@email#1#2{%
 \endgroup
 \patchcmd{\titleblock@produce}
  {\frontmatter@RRAPformat}
  {\frontmatter@RRAPformat{\produce@RRAP{*#1\href{mailto:#2}{#2}}}\frontmatter@RRAPformat}
  {}{}
}%
\makeatother
\DeclareUnicodeCharacter{2212}{-}
\newcommand{\rep}[2]{{#1}}{}
\newcommand{\repxb}[1]{\textcolor{black}{#1}}

\title{Polarizable Continuum Models and Green's Function $\bf{GW}$ Formalism: On the Dynamics of the Solvent Electrons}

\author{Ivan Duchemin}
\affiliation{Univ. Grenoble Alpes, CEA, IRIG-MEM-L\_Sim, 38054 Grenoble, France}
\email{ivan.duchemin@cea.fr}
\author{David Amblard}
\affiliation{Univ. Grenoble Alpes, CNRS, Inst NEEL, F-38042 Grenoble, France}
\author{Xavier Blase}
\affiliation{Univ. Grenoble Alpes, CNRS, Inst NEEL, F-38042 Grenoble, France}
\email{xavier.blase@neel.cnrs.fr}

\SectionNumbersOn

\begin{document}	


\begin{tocentry}
\begin{center}
\includegraphics[width=8.0cm]{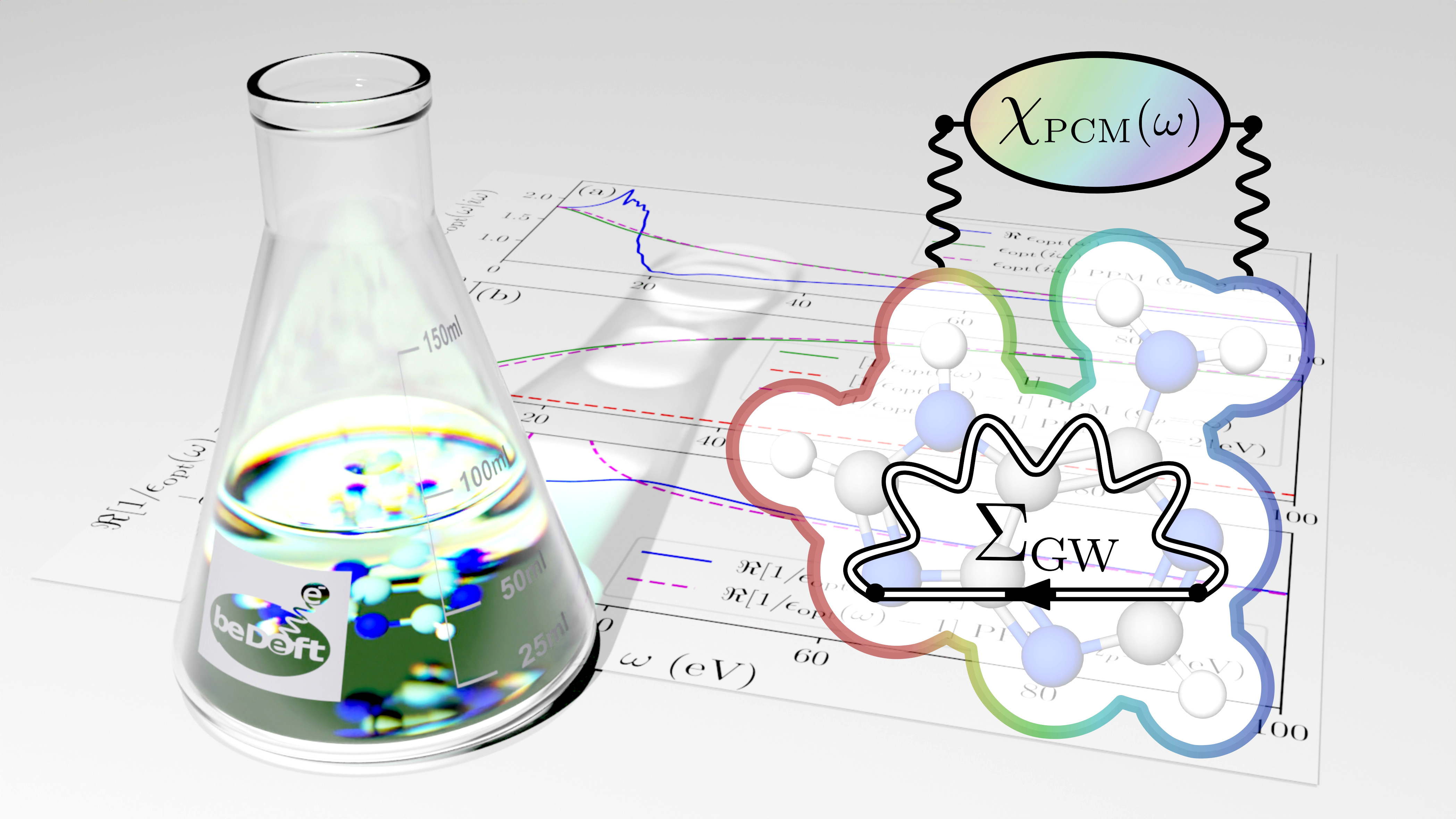}
\end{center}
\end{tocentry}

\begin{abstract}
\rep{The many-body $GW$ formalism, for the calculation of ionization potentials or electronic affinities, relies on the frequency-dependent dielectric function built from the electronic degrees of freedom.  Considering the case of water as a solvent treated within the polarizable continuum model, we   explore the impact    of restricting the full frequency-dependence of the solvent electronic dielectric response  to a frequency-independent $(\epsilon_\infty)$ optical dielectric constant. For solutes presenting small to large highest-occupied to lowest-unoccupied molecular orbital energy gaps, we show that such a restriction induces errors no larger than a few percent on the energy level shifts from the gas  to the solvated phase.   We further introduce a remarkably accurate single-pole model for mimicking the effect of the full frequency dependence of the water dielectric function in the visible-UV range. This allows a fully dynamical embedded $GW$ calculation with the only knowledge of the cavity reaction field calculated for the  $\epsilon_\infty$  optical dielectric constant.} 
{ Standard polarizable continuum models (PCM)  account for the low frequency limit of the optical dielectric constant. This relies on the assumption that the solvent electrons react instantaneously to the solute electronic excitations, treating thus the solvent electrons in the adiabatic limit. Describing the solute at the  many-body $GW$ level, we explore the impact on the solute electronic energy levels of switching the full frequency-dependence of the solvent optical dielectric response. Considering the case of water as a solvent, we show that the adiabatic limit  introduces errors smaller than a few percent on the polarization energies, even for solutes presenting large highest-occupied to lowest-unoccupied molecular orbital energy gaps. We further introduce a remarkably accurate single-pole model for the frequency-dependent water dielectric constant in the optical range. This allows a fully dynamical calculation with the only knowledge of the cavity reaction fields calculated for the low-frequency optical dielectric constant, as in standard PCM calculations.  }
\end{abstract}

\maketitle

\section{Introduction}
\label{sec:intro}

The renormalization of electronic excitations by a polarizable environment, such as an electrode, an organic crystal or a solvent, plays an important role in many fields pertaining to physics, chemistry, and biology. This is particularly true in the case of charged excitations, as described by direct or inverse photoemission, where the reaction field from the environment can stabilize the added charge by several electronvolts. The stabilization energy, labeled a polarization energy, depends on the environment dielectric properties. The complexity of the environment, in many applications of interest, initiated hybrid theoretical strategies, merging the quantum mechanical treatment of a central active subsystem, with a simplified description of the environment response properties described as a continuum or discrete (atomistic) medium. \cite{WARSHEL1976,Miertus_1981}  

Originating in simple image models for the description of a charge facing a dielectric surface or located inside a dielectric cavity, \cite{Born_1920,Jackson}   the polarizable continuum model (PCM) in its various implementations   has emerged as a very successful approach, \cite{Miertus_1981,Klamt_1993,Cammi_1994,Cammi_1995,Tomasi_1994,Cances_1997,Barone_1998,Tomasi_2005}  in particular in the description of solvated species for which more accurate but expensive atomistic simulations require averaging over molecular dynamics trajectories. \rep{In the PCM approach, the solute is placed in a cavity carved into an homogeneous medium characterized by its macroscopic dielectric function $\epsilon(\omega)$ that depends on general principles on the excitation frequency. Of importance for the following discussions, two specific solvent dielectric constants can be introduced when it comes to study fast electronic excitations,}{ Standard PCM calculations rely on the use of two different dielectric constants} namely the static and optical dielectric constants labeled $\epsilon_0$ and $\epsilon_{\infty}$, respectively. The first one (e.g., $\epsilon_0$=78.35 for water) includes the slow ionic motions, while the second ($\epsilon_{\infty}$=1.78 for water) only accounts for the fast electronic degrees of freedom. 
The use of  \rep{different  dielectric constants in different frequency domains}{ these two dielectric constants} leads to the so-called non-equilibrium formalisms.\cite{Cammi_1995} In practice, \rep{within this very simplified description of the   frequency dependence of the dielectric constant,}{}  $\epsilon_0$ is employed to renormalize the ground-state properties (charge density, one-body molecular orbitals, etc.) while $\epsilon_{\infty}$ is adopted for the solvent reaction to the fast electronic excitations from the ground-state to an excited state. 

\repxb{The solvent is chosen to be mostly transparent (non-absorbing) in the energy range of the solute lowest electronic excitations of interest. As such, the poles of the solvent electronic susceptibility are expected to be located at significantly higher energy. This may certainly justify the use of a frequency-independent $\epsilon_{\infty}$ optical dielectric constant defined from the static limit of the solvent electronic susceptibility. The subscript $\infty$ really points to a frequency range significantly larger than the typical ionic frequencies, even though smaller than the energy of the solvent electronic susceptibility poles.} \rep{Restricting the PCM dielectric response to  the $\epsilon_0$ and $\epsilon_{\infty}$ dielectric constants can be improved by considering the full frequency-dependent $\epsilon(\omega)$ macroscopic dielectric constant for frequencies  spanning the typical ionic to electronic time scales. This was introduced within a time-dependent non-equilibrium dielectric response PCM approach.
\cite{Mennucci_2005,Xiaosong_2012_JCPL,Xiaosong_2012_JCPA,Cammi_2015_EOM-PCM,Xiaosong_2015_TDPCM,Corni_2019_RT-TDDFT,Corni_2020_NP_RTTDDFT}  Concerning the solvent/solute problem,
the frequency dependence often relied on the   Debye  relaxation model for the solvent degrees of freedom dynamics, interpolating between $\epsilon_0$ and $\epsilon_{\infty}$.   \cite{Mennucci_2005,Cammi_2015_EOM-PCM,Xiaosong_2015_TDPCM,Corni_2019_RT-TDDFT} 
As such, the high-frequency limit of the dielectric response was set to $\epsilon_{\infty}$. In practice, and as shown in Fig.~\ref{fig:epsilon}(a) (blue line) for water as the solvent, the macroscopic optical dielectric function,  the square of the refractive index, shows significant variations with frequency in the visible-UV range, and $\epsilon_{\infty}$=1.78 for water corresponds to the low frequency limit response from the electronic degrees of freedom. In this spirit, a fit to experimental measurements of the frequency-dependent dielectric function in the visible-UV range was exploited for the time-dependent study of the reaction field close to metallic nanoparticles treated as a continuous polarizable medium. \cite{Corni_2020_NP_RTTDDFT}   }{}
\repxb{Besides specific formulations of the frequency-dependent macroscopic dielectric function, a general PCM approach with a fully frequency-dependent  solvent response matrix  was introduced within an open quantum system theory framework. \cite{Guido_2020} }

\rep{As another issue, the comparison between the dynamics of the solvent and solute electronic degrees of freedom leads to two limiting regimes. If the low lying poles of the solute electronic susceptibility are located at much lower energy than that of the solvent, then the solvent electronic response to the solute low-lying excitations can be considered to be instantaneous. This is called the Born-Oppenheimer (BO) regime in the PCM literature. \cite{Kim_JCP_1992,Kuznetsov_1992,Marcus_1992,Guido_2020}  
The opposite regime, where the solvent electronic degrees of freedom are assumed to be slower than that of the solute, is called the self-consistent regime. Clearly, in cases where the decoupling of energy between the solute and solvent electron dynamics cannot be made, none of these regime strictly applies. In particular, the actual frequency dependence of the solvent optical dielectric response needs to be considered.}{}

Already merged with the PCM formalism, \cite{Duchemin2016,Kim2022,Clary2023} the Green's function $GW$ many-body perturbation theory, \cite{Hed65} where $G$   stands for the one-body Green's function and $W$ for the screened Coulomb potential,   has   recently gained much popularity in quantum chemistry.   Significant efforts have been devoted to benchmark its accuracy for molecular systems as compared to quantum chemistry techniques such as coupled-cluster approaches, \cite{Rostgaard2010,Blase2011,Marom2012,Bruneval2013,vanSetten2015,Krause2015,Knight2016,Kaplan2016,Rangel2016,Caruso2016,Maggio2017,Govoni2018,Forster2021} together with much progress in developing low-scaling implementations.
\cite{Rojas1995,Foerster2011,Neuhauser2014,Liu2016,Vlcek2017,Vlcek2018,Wilhelm2018,Forster2020,Kim2020,Kupetov2020,Gao2020,Wilhelm2021,Duchemin2021,Forster2023,Scott2023}
Targeting the calculation of electronic energy levels, \rep{including ionization potentials and electronic affinities,}{} properly defined as charging energies, the $GW$ operator can be described as a non-local and dynamical self-energy term that includes exchange and correlation effects. In particular, the screened Coulomb potential $W({\bf r},{\bf r}'; \omega)$ is fully dynamical, accounting for the frequency dependence in the visible-UV range of the dielectric function $\epsilon({\bf r},{\bf r}'; \omega)$. As such, the $GW$ formalism offers an obvious path to explore fully dynamical models of polarizable environments, with in particular a macroscopic dielectric function \rep{originating from electronic degrees of freedom}{ $\epsilon_{\infty}$} that is not restricted to   \rep{a single $\epsilon_{\infty}$ dielectric constant}{}, but is instead allowed to depend on the excitation frequency in the visible-UV range. 

\rep{Replacing the fully dynamical optical dielectric function of Fig.~\ref{fig:epsilon}(a) by the  frequency-independent $\epsilon_{\infty}$ constant, amounts to pushing the poles of the solvent electronic susceptibility to infinity. This leads to the BO regime where the solvent electronic degrees of freedom are assumed to have a much faster dynamics than that of the solute.  Such an analysis is consistent with the conclusions provided by Guido and coworkers in Ref.~\citenum{Guido_2020} concerning the early merging \cite{Duchemin2016} of the $GW$ formalism with a PCM optical dielectric function fixed to the $\epsilon_{\infty}$ constant.}{ In the standard non-equilibrium PCM approach, the restriction of the environment optical dielectric constant to its low frequency $\epsilon_{\infty}$ limit relies implicitly on the assumption that the environment electronic degrees of freedom react instantaneously to an electronic excitation on the solute. }
Such \rep{an approximation of a much faster solvent electrons dynamics}{ an adiabatic  approximation for the solvent} is supposed to be valid when the solvent transition energies from occupied to unoccupied states are located at significantly higher energy than the solute analogs. 
\rep{}{The limits of this assumption were first explored on the basis of model solvents and solutes. [Painelli2020] }
\rep{Reintroducing the full frequency-dependence of the solvent optical dielectric response, as provided in Fig.~\ref{fig:epsilon} for water,  allows to tackle systems where solute and solvent electron dynamics are not decoupled, including situations where the poles of the solute electronic susceptibility may be located at higher energies than that of the solvent.}{}
Recently, a fully dynamical PCM has been introduced and integrated with the $GW$ formalism, \cite{Clary2023} but no comparison was made to \rep{an approach where the frequency-dependent optical dielectric function is restricted to the $\epsilon_{\infty}$ constant.}{ its standard static version.}

In a recent work,\cite{Amblard2024} the authors merged the $GW$ formalism with a dynamical treatment of the environment degrees of freedom in a fully \textit{ab initio} QM/QM'   scheme.  These developments allowed in particular to scrutinize the accuracy of the \rep{instantaneous environment electrons response}{ adiabatic} approximation. \cite{noteadiabatic} Considering  a molecular system immersed in its parent molecular crystal, that is a situation where there is no decoupling of energy between the ``solute'' and ``solvent'' electronic degrees of freedom, the error associated with treating the environment in \rep{the BO}{ its adiabatic}   limit was shown to induce errors no larger than 10$\%$ for the polarization energy associated with frontier orbitals or the energy gap. Such an \textit{ab initio} and dynamical treatment of the environment stands as a generalization of previous QM/MM$_{\text{pol}}$ implementations \cite{Baumeier2014,Duchemin2016,Li2016,Li2018,Wehner2018,Tirimbo2020} based on semi-empirical and low-frequency descriptions  of the environment dielectric properties in the optical range. \cite{Thole1981,D_Avino_2016,mennucci-MMpol-2020}    


In the present study, we explore explicitly the effect of considering the full frequency dependence of the optical dielectric constant of  water as a solvent described by the PCM. More explicitly, we calculate  the $GW$ ionization potential and electronic affinities of solvated molecules, switching on and off the frequency-dependence of the optical macroscopic dielectric constant of water. For a large set of molecules showing very different highest-occupied to lowest-unoccupied molecular orbital (HOMO-LUMO) energy gaps, we show that treating the solvent in \rep{the BO}{ its adiabatic}  limit does not induce errors larger than a few percent on the polarization energy. We further introduce a simple pole-model for the solvent electronic dielectric response that accurately reproduces the effect of considering the full dynamics of the solvent electronic degrees of freedom, while only requiring the calculation of the PCM reaction field  \rep{associated with the $\epsilon_{\infty}$ optical dielectric constant.}{ in the low frequency limit as in standard PCM calculations.}

\section{Theory }  

We start this theory section with a brief outline of the specific $GW$ features and flavors that will be used hereafter to support strategies for merging with a polarizable environment. Broader and extensive descriptions of the Hedin's formalism can be found elsewhere, as we refer the reader to either seminal articles \cite{Hed65,Str80,Hyb86,God88,Lin88,Farid88} or more recent books or reviews.\cite{Ary98,Farid99,Oni02,Pin13,Leng2016,ReiningBook,Gol19rev}  We will keep our demonstration to the quantities relevant to the present work.

\subsection{The $GW$ formalism}

A standard $GW$ calculation starts from an input time ordered Green's function $G({\bf r},{\bf r}'; \omega)$, built upon $\{\varepsilon_n,\phi_n\}$ Kohn-Sham eigenstates, and constructs the dynamical $GW$ exchange-correlation self-energy operator defined as
\begin{equation}
 \Sigma({\bf r},{\bf r}'; E) = \frac{i}{2\pi} \int \dd\omega  \; e^{i \eta \omega} G({\bf r},{\bf r}'; E+\omega)  W({\bf r},{\bf r}'; \omega),
 \label{eqn:sigma}
\end{equation}
with $\eta$ a positive infinitesimal. The dynamically screened Coulomb potential $W({\bf r},{\bf r}'; \omega)$ comes instead of  the bare Coulomb potential $v({\bf r},{\bf r}')$ that would lead to the  Hartree-Fock exact exchange $Gv$ operator. 
As the self-energy is dynamical, performing a $GW$ calculation in a polarizable or dielectric environment relies straightforwardly on the relation between $W({\bf r},{\bf r}'; \omega)$ and the inverse dielectric function ${\epsilon}^{-1}({\bf r},{\bf r}'; \omega)$: 
\begin{equation}
  \begin{split}
    W({\bf r},{\bf r}'; \omega) & = \int \dd{\bf r}'' \; {\epsilon}^{-1}({\bf r},{\bf r}''; \omega)\, v({\bf r}'',{\bf r}') \\
     & = {\epsilon}^{-1}(\omega) \cdot v ,
    \label{eqn:W_eps}
  \end{split}
\end{equation}
where we introduce the $\cdot$ product to indicate composition over the space variables of two adjacent \rep{linear}{} operators, \rep{i.e. the result of their successive application as defined by the matrix product of their matrix representations}{}. 
Since the $GW$ formalism tackles the study of electronic correlation effects, we emphasize that the dielectric function ${\epsilon}^{-1}(\omega)$ considered here shall only account for the electronic degrees of freedom, justifying below the notation $\epsilon_{\text{opt}}$ where ``{o}pt'' stands for optical. In contrast, the effect of the slow ionic degrees of freedom was accounted for at the Density Functional Theory (DFT) level, e.g., generating initial DFT@PCM Kohn-Sham states with a dielectric constant of $\epsilon_0$ = 78.35 for water as the solvent. \cite{pcmnotations}  

Alternatively, one can define $W({\bf r},{\bf r}'; \omega)$ through the electronic susceptibility ${\chi}({\bf r},{\bf r}'; \omega)$:
\begin{equation}
  \begin{split}
    W(\omega) = & \; v + v \cdot {\chi}(\omega) \cdot v  ,
     \label{eqn:W_chi}
  \end{split}
\end{equation}
which is itself related to the free susceptibility $\chi_0({\bf r},{\bf r}'; \omega)$:
\begin{equation}
  \begin{split}
    \chi(\omega) = & \; \chi_0(\omega) 
    + \chi_0(\omega) \cdot v \cdot {\chi}(\omega),
  \end{split}
\label{eqn:chi_chi0}
\end{equation}
adopting the Random Phase Approximation (RPA). All the above quantities are dynamical, and require in particular the knowledge of the full dynamics of ${\epsilon}^{-1}_{\text{opt}}({\bf r},{\bf r}'; \omega)$. As such, merging with low frequency limit models such as PCM cannot be made straightforwardly as taking $\epsilon_{\infty} = \epsilon_{\text{opt}}(\omega \rightarrow 0)$ for the environment would lead to the wrong high frequency limit for $W({\bf r},{\bf r}'; \omega)$. 
\\

The difficulties brought by the frequency dependence within the $GW$ self-energy of eq \ref{eqn:sigma} were circumvented by Hedin\cite{Hed65} through the introduction of the  so-called Coulomb-hole (COH) plus screened-exchange (SEX) static approximation:
\begin{align}
\Sigma^{\text{SEX}}({\bf r},{\bf r}') &= - \sum_i^{\text{occ}}  \phi_i({\bf r}) \,\phi^{*}_i({\bf r}') \, W({\bf r},{\bf r}';\omega=0) \label{SEX}\\
\Sigma^{\text{COH}}({\bf r},{\bf r}') &= \frac{1}{2} \,\sum_n \phi_n({\bf r}) \,\phi^{*}_n({\bf r}') \left[ W({\bf r},{\bf r}';\omega=0)-v({\bf r},{\bf r}') \right], \label{COH}
\end{align}
requiring only the low-frequency $W(\omega \rightarrow 0)$ limit of the screened Coulomb potential, or equivalently of the susceptibility $\chi(\omega \rightarrow 0)$ . 
An elegant and simple way to recover the static COHSEX approximation is to consider that the system's electron density reacts immediately to any perturbation within the $GW$ equations response functions \ref{eqn:W_chi} and \ref{eqn:chi_chi0}.\cite{Amblard2024} In this case, the dynamical response of the system may be described by a simple pole model tailored to reproduce this low frequency limit:    
\begin{equation}
\begin{split}
    \chi^{}_{\Omega_p}({\bf r},{\bf r}' ;\omega) &= \chi({\bf r},{\bf r}'; \omega=0)    
          \times f(\omega; \Omega_p)
\end{split}
\label{eqn:ximodel}
\end{equation}
with
\begin{equation}
f(\omega; \Omega_p) =
     \frac{\Omega_p}{2}\left[  \frac{1}{\omega + \Omega_p-i\eta}- \frac{1}{\omega - \Omega_p+i\eta}  \right], 
     \label{eqn:fpole}
\end{equation}
and $\Omega_p$ a unique pole energy for simplicity. Such an expression has the correct static limit and time-ordering structure in the energy plane. \rep{}{The adiabatic limit, i.e.,} The static COHSEX self energy \ref{SEX}+\ref{COH}, is recovered by taking the pole energy $\Omega_p$ to infinity \textit{after} performing the energy integration in eq \ref{eqn:sigma}. 
We will exploit here below the same strategy in order to merge the standard PCM model, described by a single $\epsilon_{\text{opt}}(\omega \rightarrow 0)$ dielectric constant, within a fully dynamical $GW$ formalism for the central subsystem. We will also further explore the possibility to efficiently account for the solvent dynamics through such a simple pole model but with a finite $\Omega_p$ pole energy.

\subsection{Embedding the $GW$ equations}

\subsubsection{ General considerations }

We start with a partition of the system into the solute $\sys$ and the solvent (i.e., environment) $\env$, and assume independent  contributions of the solute and solvent free susceptibilities $\chi_0^\sys$ and $\chi_0^\env$ to the free susceptibility $\chi^{}_0$ of the full system. This assumption is fully justified in the case of non-overlapping wave-functions between subsystems $\sys$ and $\env$. In such case, the screened Coulomb potential can be obtained through:
\begin{equation}
\begin{split}
  W^{-1}(\omega) & = v^{-1} - \big(\,\chi_0^\sys(\omega) + \chi_0^\env(\omega)\,\big) \\
                 & = \tilde{v}^{-1}(\omega) - \chi_0^\sys(\omega),
\end{split}
\label{eq:W_full}
\end{equation}
with the frequency dependent screened Coulomb interactions $\tilde{v}(\omega)$ defined through the Woodbury identity:
\begin{equation}
\begin{split}
  \tilde{v} & = \big(\, v^{-1} - \chi_0^\env(\omega) \,\big)^{-1} \\
            & = v + v \cdot \chi^\env(\omega) \cdot v, \\
\end{split}
\end{equation}
and $\chi^\env(\omega)$ is the interacting susceptibility of the environment, or solvent, \textit{in the absence of the solute}
\begin{equation}
\begin{split}
  \chi^\env(\omega)^{-1} & = \chi_0^\env(\omega)^{-1}-v. \\
\end{split}
\end{equation}
For clarity in the following equations, we will write
\begin{equation}
\begin{split}
  v_{\text{reac}}(\omega) & = v \cdot \chi^\env(\omega) \cdot v
\end{split}
\end{equation}
the reaction potential of the solvent induced by the solute: qualitatively, a change of the solute charge density induces a change in the solvent charge density proportional to $\chi^\env \cdot v$ that in return exerts a reaction potential on the solute via $v$.  
Finally, the solution to eq \ref{eq:W_full} can be decomposed in 
\begin{equation}
    W(\omega) ={\tilde v }(\omega)+{\tilde v }(\omega)\cdot{\tilde \chi}(\omega)\cdot{\tilde v }(\omega) \\
\end{equation}
and
\begin{equation}
    {\tilde \chi}(\omega) = \chi_0^\sys(\omega)+ \chi_0^\sys(\omega)\cdot{\tilde v }(\omega)\cdot{\tilde \chi}(\omega), \label{eqn:tildexi}
\end{equation}
which follows the ingredients of a standard isolated $GW$ calculation, but where ${\tilde \chi}(\omega)$ is the interacting susceptibility of the solute obtained with the frequency dependent Coulomb interactions that have been screened through the solvent reaction potential.

\subsubsection{Dynamical PCM models}

The dynamical cavity reaction field $v_{\text{reac}}(\omega)$ can be seen as the result of a frequency-dependent PCM model associated with the frequency-dependent $\epsilon_{\text{opt}}(\omega)$ solvent optical dielectric constant. The PCM formalism can thus be straightforwardly generalized to a fully dynamical scheme, recalculating the cavity reaction fields for each frequency involved in the energy-quadrature used to evaluate the self-energy (see eq \ref{eqn:sigma}).  In our approach combining the contour-deformation scheme with the analytic continuation of the screened Coulomb potential  (see Technical details section \ref{tech_det}),\cite{Duchemin2020}  this means recalculating the cavity charges for  typically  twelve imaginary frequencies. Since the dielectric constant is real along the imaginary axis, the standard continuity equations on the cavity surface \cite{Cances_1997,Duchemin2016} can be applied without modifications. 

\begin{figure}
  \includegraphics[width=8.6cm]{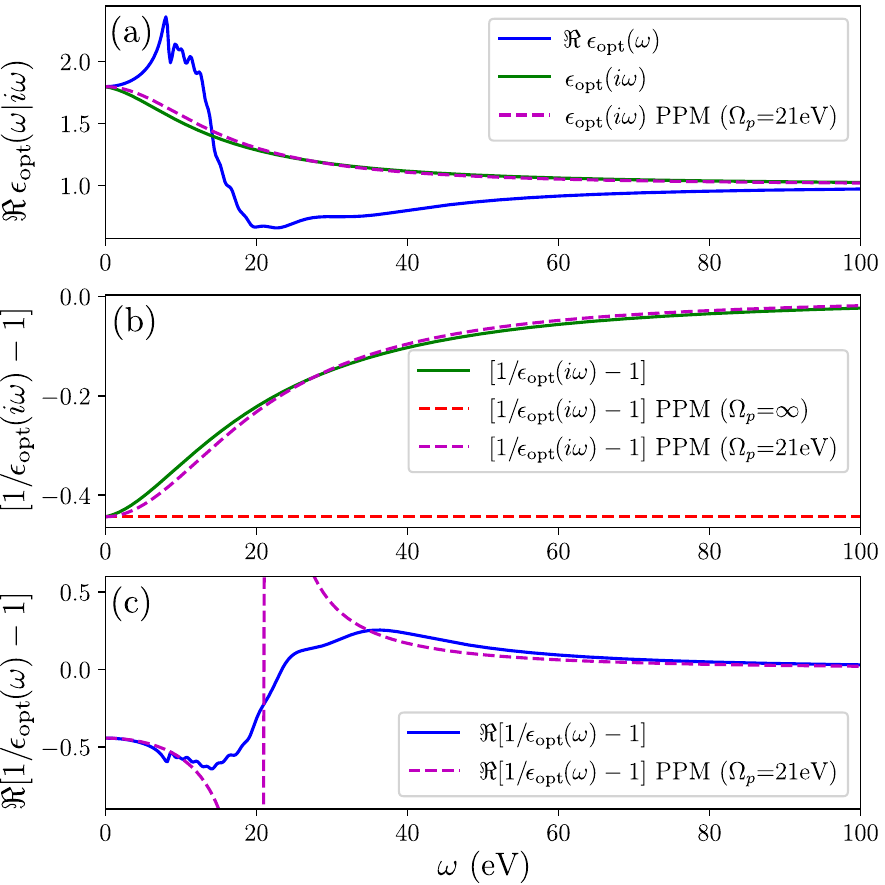} 
  \caption{ (a) Real-part of the macroscopic optical dielectric constant of water ($\epsilon_{\text{opt}}$) along the real (blue line) and imaginary  (green line) frequency  axes [from Ref.~\citenum{Fiedler2020}]. \rep{The $\omega=0$ limit allows to recover the $\epsilon_{\infty}=1.78$ optical dielectric constant.}{}
 The dashed purple line represents the dielectric constant at imaginary frequencies, as obtained with the one-pole plasmon-pole model (PPM) described in the main text (eq \ref{eqn:xipcm_1}).  (b) Plot of $[ 1/\epsilon_{\text{opt}}(i\omega)-1]$  from Ref.~\citenum{Fiedler2020} (green line) and corresponding single-pole functional fit with a pole  located at $\Omega_p$=21 eV (dashed purple line). The red dashed line represents the \rep{limit case where $\Omega_p$ is}{ static (adiabatic) approach with $\Omega_p$} pushed to infinity. (c) Real-part of $[1/\epsilon_{\text{opt}}(\omega)-1]$   from Ref.~\citenum{Fiedler2020} (blue line) and using the single-pole functional fit with a pole at 21 eV and a 0.5 eV broadening (dashed purple line). }
  \label{fig:epsilon}
\end{figure}

The  frequency dependence of the water macroscopic optical dielectric constant was recently parametrized under the form of damped harmonic oscillators reproducing faithfully available experimental data.\cite{Fiedler2020}  The resulting frequency-dependent optical dielectric constant $\epsilon_{\text{opt}}(\omega)$ is  represented in Fig.~\ref{fig:epsilon}(a) along the real-axis (blue line) and along the imaginary-axis (green line).  We further plot the related
$[ 1 / \epsilon_{\text{opt}}(\omega) -1 ]$ along the imaginary axis (green line, Fig.~\ref{fig:epsilon}(b)) and along the real axis 
(blue line, Fig.~\ref{fig:epsilon}(c)). As inferred from eqs \ref{eqn:W_eps} and \ref{eqn:W_chi}, this quantity is more closely related to the susceptibility $\chi$ entering the reaction field:
\begin{equation}
  \epsilon_{\text{opt}}(\omega)^{-1} = 1 + v \cdot 
\chi(\omega).
\end{equation}
In particular, $\epsilon_{\text{opt}}(\omega)^{-1}$ and $\chi(\omega)$ share the same pole structure. 

The fit of Ref.~\citenum{Fiedler2020} relies on damped oscillators. This is a difficulty since the resulting susceptibility, or related $\epsilon_{\text{opt}}(\omega)^{-1}$ inverse macroscopic dielectric constant,    does not offer the proper symmetry along the real and imaginary-frequency  axes. We bypass this problem by showing in Fig.~\ref{fig:epsilon} that a  simple single-pole model with proper symmetry:
\begin{myleftbar}
{\bf \emph{ dynamical PCM($\omega$):}}
\begin{equation}
\begin{split}
    \epsilon_{\text{opt}}(\omega)^{-1} = 1  +   \; \Big(\,\epsilon_{\infty}^{-1}-1\,\Big) \times f(\omega; \Omega_p)  
\end{split}
\label{eqn:xipcm_1}
\end{equation}
\end{myleftbar}
\noindent offers an accurate approximation to the full fit along the imaginary axis, fixing the pole energy at $\Omega_p$=21~eV.  The resulting $[ 1 / \epsilon_{\text{opt}}(\omega)-1]$ function (dashed purple line, Fig.~\ref{fig:epsilon}(b)) follows closely the original fit along the imaginary axis. A similar agreement is found for the related $\epsilon_{\text{opt}}(\omega)$ function
(dashed purple line, Fig.~\ref{fig:epsilon}(a)). Even though not needed for our analytic continuation scheme applied to frontier orbitals, we show in Fig.~\ref{fig:epsilon}(c) that the single-pole fit is accurate at low energy along the real-energy axis. As shown in the Supporting Information (SI, Figs.~S1 and S2), a more accurate fit, allowing for several poles, does not change the resulting polarization energies by more than a very few meV. Such a scheme, and the related polarization energies, represent our reference dynamical PCM calculations and will be labeled $\mathrm{PCM(\omega)}$.
\\

Fully dynamical PCM calculations, requiring to recalculate the cavity reaction fields at several frequencies, increase the complexity and computational cost as compared to the \rep{BO limit (fast solvent electron dynamics) where only the $\epsilon_{\infty}$ optical dielectric constant needs to be considered.}{ standard adiabatic PCM.}  We now introduce a seemingly severe approximation by modeling the dynamics of the non-local susceptibility  as follows:
\begin{myleftbar}
{\bf \emph{ plasmon-pole model solvent ($\mathrm{PCM_{PP}}$):}}
\begin{equation}
\begin{split}
    \chi^\env_{\Omega_p}({\bf r},{\bf r}';\omega) & = \chi^\env_{\text{PCM}}({\bf r},{\bf r}';\omega=0)  \times f(\omega; \Omega_p). \\ 
\end{split}
\label{eqn:xipcm_2}
\end{equation}
\end{myleftbar}
While the fit of the experimental   macroscopic $\epsilon_{\text{opt}}(\omega)^{-1}$  function  by a single-pole function (eq \ref{eqn:xipcm_1}) represents just a properly symmetrized representation of a frequency-dependent scalar, the latter approximation is much more questionable since it introduces a decoupling between the real-space and frequency degrees of freedom. On formal grounds,  the solvent susceptibility reads:
\begin{align*}
\chi^{\env}( {\bf r},{\bf r}' ; \omega) = \sum_n^{\text{poles}}       \left[  \frac{A_n({\bf r},{\bf r}')}{\omega - \Omega_n+i\eta} - \frac{A_n({\bf r},{\bf r}')}{\omega + \Omega_n-i\eta}    \right] , 
\end{align*}
with $A_n({\bf r},{\bf r}') = \rho_n({\bf r}) \rho_n({\bf r}')$ and where the $\Omega_n$ are the RPA transition energies and $\rho_n({\bf r})$ the corresponding transition densities. Clearly, the spatial and frequency degrees of freedom are entangled. The decoupling of the spatial and frequency degrees of freedom for the solvent susceptibility extends to the reaction field
$v_{\text{reac}}(\omega) = v \cdot \chi^{E}(\omega) \cdot v$. As such, the reaction field at finite frequency can be straightforwardly obtained by multiplying the static reaction field by the dynamical factor $f(\omega, \Omega_P)$.
This is a dramatic simplification since only the low-frequency limit of the reaction fields needs to be calculated as in a standard PCM calculation. 
A similar approach was recently proposed to improve on the static COHSEX approximation for calculating the IP and EA of molecular systems. \cite{Tyagi_2024} Our approach is less ambitious as we only intend to capture the dynamics of the environment susceptibility.  The solute electrons correlation energy is in our case treated at the fully dynamical $GW$ level.

The validation of this plasmon-pole  approximation for the solvent susceptibility, as compared to the dynamical PCM approach where the reaction fields are explicitly recalculated for each $\epsilon_{\text{opt}}(\omega)$ value, represents one of the main goal of this paper. 
This approach will be labeled $\mathrm{PCM_{PP}}$ in what follows, where PP stands for plasmon-pole.

\subsubsection{  The static COHSEX PCM model: $\Omega_p\to\infty$  }

To connect to  \rep{}{standard} PCM calculations that neglect the frequency dependence of the optical dielectric constant, and in order to explore the consequences of such an approximation as compared to the fully dynamical PCM, we now turn to $GW$@PCM calculations in the \rep{BO }{adiabatic} limit, namely assuming that the solvent degrees of freedom react instantaneously to any excitation on the solute. As shown recently in a fully \textit{ab initio} QM/QM' implementation of embedded $GW$ calculations, \cite{Amblard2024} this can be very simply obtained from eq \ref{eqn:xipcm_2} by bringing the $\Omega_p$ frequency to infinity \textit{after performing the frequency  integration} that defines the $GW$ self-energy (eq \ref{eqn:sigma}). This leads to the so-called static Coulomb-Hole plus Screened-Exchange (COHSEX) decomposition of the reaction field as briefly described now. More details can be found in Ref.~\citenum{Amblard2024}.

We focus on the correlation-only $\Sigma_\text{C}(E)$ self-energy, leaving aside the bare exchange contribution, with:  
\begin{equation}
 \Sigma_\text{C}(E) = \frac{i}{2\pi} \int \dd\omega \; e^{i \eta \omega} G(E+\omega)    \Hp \big[W(\omega)-v\big],
 \label{eqn:sigma_c}
\end{equation}
where the space variables are omitted for the sake of compactness. This equation can be rewritten in terms of the embedded Coulomb interactions and reaction field as
\begin{equation}
\begin{split}
 \Sigma_\text{C}(E) &  =  \frac{i}{2\pi} \int \dd\omega \; e^{i \eta \omega} G(E+\omega) \Hp  \big[W(\omega)-{\tilde v(\omega)}\big] \\
&  + \frac{i}{2\pi} \int \dd\omega \; e^{i \eta \omega} G(E+\omega) \Hp  v_{\text{reac}}(\omega).  
 \label{eqn:sigma_c2}
\end{split}
\end{equation}
It turns out that, beyond its similarity with the usual formulation \ref{eqn:sigma_c}, the first term of eq \ref{eqn:sigma_c2} brings a contribution to the self energy that can be computed using only the low frequency limit of ${\tilde v}(\omega)$. This is thanks to the contribution of $W(\omega)-{\tilde v}(\omega)$ poles coming from the solvent dynamics vanishing as $\Omega_p\to\infty$. Thanks to that, and as detailed in appendix \ref{appendix_limits}, we can write:  
\begin{equation}
\begin{split}
   \lim_{\Omega_p\to\infty  } & \frac{i}{2\pi} \int \dd\omega \; e^{i \eta \omega} G(E+\omega) \Hp  \big[W(\omega)-{\tilde v(\omega)}\big] \\
 = & \frac{i}{2\pi} \int \dd\omega \; e^{i \eta \omega} G(E+\omega) \Hp  \big[{\widetilde W}(\omega)-{\tilde v(0)}\big]
 \label{eqn:sigma_c2_1}
\end{split}
\end{equation}
with 
\begin{equation}
{\widetilde W}(\omega)^{-1} = {\tilde v}(0)^{-1} - \chi_0^\sys(\omega). 
\end{equation}
Remains thus only the contribution from the finite frequency poles of the solute, that appears as re-normalized within the static limit of the solvent.  
The direct contribution of the poles of the environment response comes from the second term of eq \ref{eqn:sigma_c2}:  
\begin{equation}
\begin{split}
\frac{i}{2\pi} \int & \dd\omega \; e^{i \eta \omega} G({\bf r},{\bf r'};E+\omega)  \,  v_{\text{reac}}({\bf r},{\bf r'};\omega) \phantom{\sum_k} \\
= & - \phantom{\frac{1}{2}} \sum_i^{\text{occ}} \phi_i({\bf r})\phi^*_i({\bf r}') v_{\text{reac}}({\bf r},{\bf r}'; \omega=0)   \\ 
  & +\frac{1}{2} \sum_n \phi_n({\bf r})\phi^*_n({\bf r}') v_{\text{reac}}({\bf r},{\bf r}'; \omega=0)
 \label{eqn:sigma_c2_2}
\end{split}
\end{equation}
where $(i)$ runs over occupied states only. The first line takes the form of a screened-exchange-like contribution to the reaction field, the second one is the analog of the Coulomb-hole (COH) contribution. Using directly the static reaction field ${v_{\text{reac}}(\omega=0)}$ without the pole structure coming from $f(\omega; \Omega_p )$ leads to the SEX contribution only, leaving aside the COH term. As shown below, this second term is crucial when looking at the polarization energy associated with individual energy levels.

We finally have an expression that differentiates between indirect and direct solvent contributions to the embedded $GW$ quasiparticle self energy:
\begin{myleftbar}
{\bf \emph{Static COHSEX PCM (\rep{BO}{adiabatic} limit):}}
\begin{equation}
\begin{split}
\Sigma_\text{C}({\bf r},{\bf r}';E) = & \frac{i}{2\pi} \int \dd\omega \; e^{i \eta \omega} G({\bf r},{\bf r}';E+\omega) \Hp  \big[{\widetilde W}({\bf r},{\bf r}';\omega)-{\tilde v({\bf r},{\bf r}';0)}\big] \phantom{\sum_k} \\
 - & \phantom{\;\frac{1}{2}} \sum_i^{\text{occ}} \phi_i({\bf r})\phi^*_i({\bf r}') \, v_{\text{reac}}({\bf r},{\bf r}'; \omega=0)   \\ 
 + & \;\frac{1}{2} \sum_n \phi_n({\bf r})\phi^*_n({\bf r}') \, v_{\text{reac}}({\bf r},{\bf r}'; \omega=0).
\end{split}
\label{eqn:COHSEX_model}
\end{equation}
\end{myleftbar}
\noindent The last two lines can be regarded as perturbative corrections to the gas phase self-energy. However, the first line integral also differs from its gas phase analog through the renormalization of the screened Coulomb and bare Coulomb potential by the static reaction field. Each term gives rise to a contribution to the polarization energy through the impact of the PCM on the correlation energy. Overall, the polarization energy reads:
\begin{align*}
P_n  = \varepsilon_n^{GW@\text{PCM}} - \varepsilon_n^{GW@\text{gas}}  
\end{align*}
where   $\varepsilon_n^{GW@\text{PCM}}$ and $\varepsilon_n^{GW@\text{gas}}$ are respectively the embedded (PCM) and gas phase quasiparticle energies. We emphasize again that in a $GW$@PCM calculation, the input Kohn-Sham eigenstates are calculated at the DFT@PCM($\epsilon_0$) level.

\subsection{ Technical details }\label{tech_det}

Our calculations are performed with the {\textsc{beDeft}} (beyond-DFT) package \cite{Duchemin2020,Duchemin2021} implementing the $GW$ and Bethe-Salpeter equation (BSE) formalisms using Gaussian basis sets and Coulomb-fitting (RI-V) resolution-of-the-identity. 
\cite{Vahtras1993,Ren2012,DucheminRI17} The correlation part of the self-energy is calculated adopting a contour-deformation scheme with an integration performed along the imaginary-frequency axis, completed by residues involving the value of the screened Coulomb potential along the real-axis. The values of $W$ along the real-energy axis can be accurately obtained by analytic continuation, a scheme much more stable than the direct analytic continuation of the full $GW$ self-energy. \cite{Duchemin2020}  $GW$ calculations are performed at the partially-self-consistent ev$GW$ level where quasiparticle energies are reinjected self-consistently in the construction of the solute Green's function $G$ and independent-particle susceptibility $\chi_0$. We adopt the def2-TZVP basis set \cite{Weigend2005} together with the corresponding  def2-TZVP-RIFIT auxiliary basis set. \cite{Weigend1998} Input Kohn-Sham eigenstates are generated using the PBE0 functional. \cite{perdew-bcp-1996,adamo-jcp-1999}

Our integral-equation-formalism (IEF) \cite{Cances_1997} PCM formulation is described in Ref.~\citenum{Duchemin2016}, where the reaction field is obtained through a double layer potential version of the formalism: the reaction field $v_{\text{reac}}$ corresponds to the potential experienced by the solute coming from the surface charges and surface dipoles induced on the cavity walls. This double layer potential has the merit of handling correctly solute charge spilling out of the cavity, which turns out to be an important point when considering virtual states. As in the standard (IEF)-PCM framework, the model depends implicitly on the solvent dielectric constant through the continuity equations at the cavity boundaries. Moreover, $v_{\text{reac}}$ is calculated in the auxiliary basis $\lbrace P \rbrace$ used to describe charge density variations in our Coulomb-fitting scheme. Namely, we calculate $v_{\text{reac}}(P,Q; \omega)$ matrix elements which can be interpreted as the action on $P$ of the reaction field associated with the surface charges and dipoles generated by the $Q$ charge density in the cavity for a given $\epsilon_\text{opt}(\omega)$ macroscopic dielectric constant.  Once the reaction-field matrix is obtained in the auxiliary basis, its action on any Kohn-Sham state is straightforward. In particular, there is no need to recalculate the reaction field components at each $GW$ iteration in the case of a self-consistent $GW$ scheme.

Input Kohn-Sham eigenstates are generated with the \textsc{Orca} package \cite{Neese2022} that implements the C-PCM version \cite{Barone_1998} of the PCM approach. Embedded DFT calculations with the water environment are thus performed using the $\epsilon_0$=78.355 dielectric constant. As such, all $GW$ calculations with the PCM environment start with the same DFT@PCM($\epsilon_0$) eigenstates. The differences between the  $GW$@$\mathrm{PCM_{COHSEX}}$   and dynamical PCM polarization energies can only stem from the way the dynamical dependence of $\epsilon_{\text{opt}}(\omega)$ is treated. 

\textbf{ List of molecules:} We study a large set of solvated molecules, starting with the adenine, cytosine, thymine, and uracil nucleobases studied in a first implementation of the $GW$@PCM formalism in the standard static PCM limit. \cite{Duchemin2016} Furthermore, as studied in the first merging of the Bethe-Salpeter equation (BSE) with the PCM, \cite{Duchemin2018} we select acrolein and indigo, the push-pull p-nitro-aniline (PNA) molecule in its planar and rotated (perpendicular)  conformation, the donor–acceptor benzene/TCNE complex, and the 4-Nitropyridine N-oxide organic probe commonly used to assess solvent polarities. Geometries are taken from Refs.~\citenum{Duchemin2016,Duchemin2018}. 
Finally, acetaldehyde, ethanol, formaldehyde, and water molecule and hexamer clusters are selected to offer a large range of  HOMO-LUMO gaps, from 5.66 eV for indigo, 13.14 eV for formaldehyde, to 16.21 eV for the water monomer at the ev$GW$@PBE0/def2-TZVP (gas phase) level. Selecting such a large spread of HOMO-LUMO gaps is expected to serve as a test of the \rep{BO approximation (faster solvent electron dynamics) }{adiabatic approximation} since such an approximation may be expected to be more robust if the gap of the solute is much smaller than that of the solvent. The water cluster geometries are taken from Refs.~\citenum{Segarra2012,Blase2016} while the acetaldehyde, ethanol, and formaldehyde molecules are relaxed at the B3LYP/def2-TZVP level.

\section{ Results and discussions }

\subsection{Accuracy of the PCM$_{\text{PP}}$ simplified dynamical scheme}

\begin{figure}[h]
  \includegraphics[width=8.6cm]{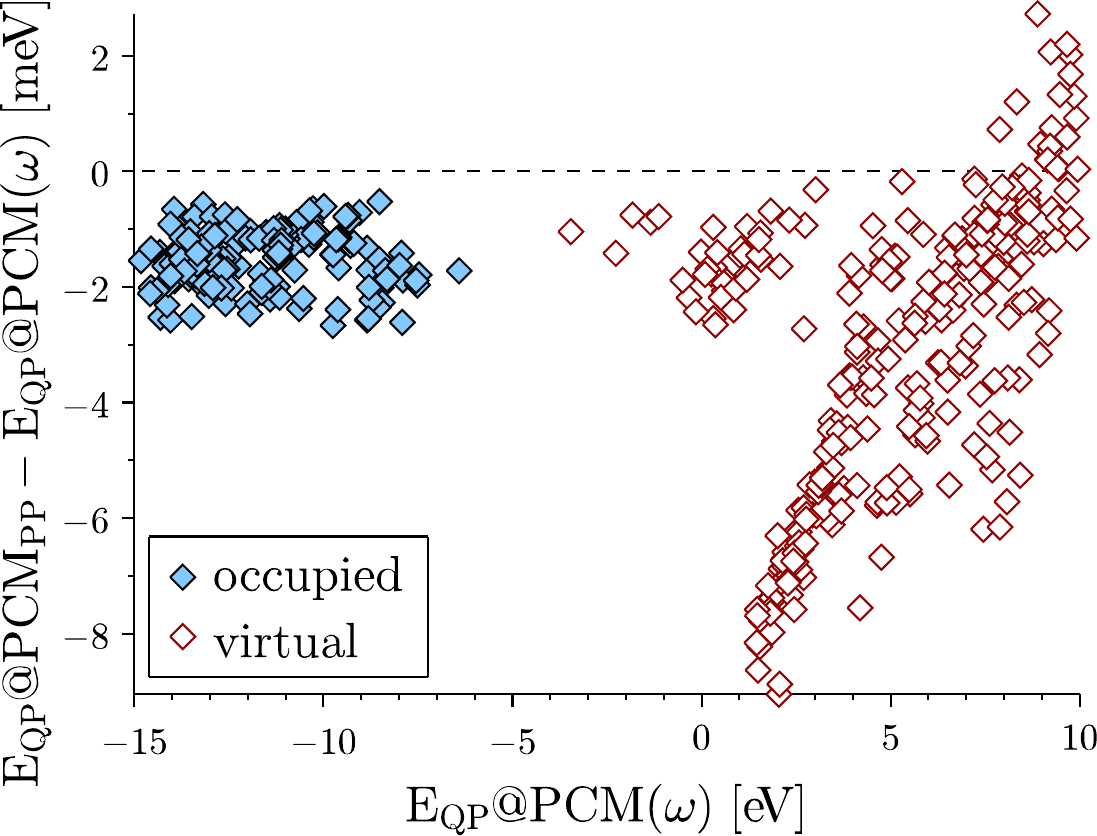} 
  \caption{ Error (meV) on the $GW$@PCM quasiparticle energies associated with the simplified (plasmon-pole) dynamical $\mathrm{PCM_{PP}}$ approach (eq \ref{eqn:xipcm_2} with $\Omega_p$=21 eV). The error is taken with respect to the fully dynamical PCM($\omega)$ approach, for which the PCM reaction fields are recalculated at each needed frequency. Errors for occupied and virtual energy levels are represented with blue-filled and red-empty diamonds, respectively. }
  \label{fig:pp_versus_w}
\end{figure}

We first compare in Fig.~\ref{fig:pp_versus_w} the error associated with the simplified plasmon-pole $GW$@$\mathrm{PCM_{PP}}$ approach where only the low-frequency reaction field is needed [see eq \ref{eqn:xipcm_2}]. The error is calculated with respect to the fully dynamical $GW$@PCM($\omega$) calculations [see eq \ref{eqn:xipcm_1}] that requires the recalculation of the reaction field at each needed frequency.

\begin{table*}
\resizebox{\columnwidth}{!}{%
\centering
\begin{tabular}{l @{\hskip 20pt} r c c c @{\hskip 20pt} r c c c }
\hline
 \multicolumn{1}{c}{}   & \multicolumn{4}{c@{\hskip 50pt}}{HOMO} & \multicolumn{4}{c@{\hskip 20pt}}{LUMO} \\
 \multicolumn{1}{c}{}   & spill. & $\mathrm{P @ PCM(\omega)}$ & $\mathrm{\Delta P @ PCM_{PP}}$  &  $\mathrm{\Delta P @  PCM_{COHSEX}}$ 
                        & spill. & $\mathrm{P @ PCM(\omega)}$ & $\mathrm{\Delta P @ PCM_{PP}}$  &  $\mathrm{\Delta P @  PCM_{COHSEX}}$ \\
\hline
Acetaldehyde        &  0.8  &  1.137 &  -0.0013  &  -0.059  &  1.6  & -1.194  &  -0.0018  &  -0.053  \\
Acroleine           &  0.7  &  1.058 &  -0.0012  &  -0.054  &  1.6  & -1.035  &  -0.0022  &  -0.056  \\
Adenine             &  0.6  &  0.741 &  -0.0018  &  -0.060  &  1.2  & -1.061  &  -0.0017  &  -0.045  \\
Benzene-TCNE        &  0.6  &  1.124 &  -0.0020  &  -0.054  &  0.5  & -0.570  &  -0.0010  &  -0.028  \\
Cytosine            &  0.7  &  0.839 &  -0.0019  &  -0.063  &  1.2  & -0.875  &  -0.0018  &  -0.059  \\
Ethanol             &  2.1  &  0.916 &  -0.0026  &  -0.087  &  2.2  & -1.361  &  -0.0019  &  -0.057  \\
Formaldehyde        &  0.8  &  1.344 &  -0.0014  &  -0.070  &  2.7  & -1.205  &  -0.0024  &  -0.070  \\
Indigo              &  0.6  &  0.677 &  -0.0017  &  -0.051  &  0.7  & -0.840  &  -0.0014  &  -0.035  \\
PNA$_{\text{perp}}$ &  0.7  &  1.400 &  -0.0017  &  -0.057  &  0.6  & -1.104  &  -0.0009  &  -0.029  \\
PNA$_{\text{plan}}$ &  0.6  &  1.120 &  -0.0019  &  -0.064  &  0.6  & -1.159  &  -0.0008  &  -0.027  \\
Probe               &  0.4  &  1.043 &  -0.0014  &  -0.052  &  0.6  & -0.773  &  -0.0008  &  -0.027  \\
Thymine             &  0.6  &  1.102 &  -0.0016  &  -0.055  &  1.1  & -0.762  &  -0.0017  &  -0.042  \\
Uracil              &  0.7  &  1.177 &  -0.0019  &  -0.061  &  1.2  & -0.766  &  -0.0018  &  -0.045  \\
\hline                                                      
\hline
mean val.           &  0.8  &  1.052 &  -0.0017  &  -0.061  &  1.2  & -0.977  &  -0.0016  &  -0.044  \\
\hline
\hline
Water               &  0.6  &  1.491 &  -0.0015  &  -0.074  &  29.2  & -1.338  &  -0.0067  &  -0.211  \\    
(H$_2$O)$_6$ book   &  0.5  &  0.787 &  -0.0012  &  -0.050  &  22.5  & -1.187  &  -0.0082  &  -0.190  \\
(H$_2$O)$_6$ cage   &  0.5  &  0.916 &  -0.0011  &  -0.053  &  20.6  & -1.106  &  -0.0076  &  -0.178  \\    
(H$_2$O)$_6$ prism  &  0.6  &  0.854 &  -0.0013  &  -0.050  &  19.1  & -1.062  &  -0.0077  &  -0.173  \\    
(H$_2$O)$_6$ ring   &  0.5  &  1.079 &  -0.0013  &  -0.054  &  24.6  & -1.445  &  -0.0086  &  -0.205  \\    
\hline                                                        
\hline
mean val.           &  0.5  &  1.025 &  -0.0013  &  -0.056  &  23.2  & -1.228  &  -0.0078  &  -0.192  \\ 
\hline
\hline
\end{tabular}
    \caption{HOMO and LUMO  PCM polarization energies  $\mathrm{P @ PCM(\omega)}$  as given by the fully dynamical $GW$@PCM($\omega$)   approach. 
The polarization energies associated with the $GW$@$\mathrm{PCM_{PP}}$ and $GW$@$\mathrm{PCM_{COHSEX}}$  are given as differences ($\mathrm{\Delta P @ PCM_{PP}}$ and $\mathrm{\Delta P @  PCM_{COHSEX}}$) with respect to the $\mathrm{P @ PCM(\omega)}$ reference. 
    All energies are given in eV. Spilling values (spill.) express the amount of each state density leaking outside of the PCM cavity (in $\%$).
    Mean values are provided for each quantity. We separate the case of water monomer and hexamers when taking the averages in relation with the large spilling of the associated LUMOs. }
    \label{tab:my_label}
} 
\end{table*}

Spanning the occupied and virtual energy levels over a large energy window for the full set of molecules, we observe that  decoupling the spatial and frequency degrees of freedom in the $\mathrm{PCM_{PP}}$ approach leads to remarkably small errors of the order of a very few meVs. Such an error is seen to increase for virtual states with positive energy (above the vacuum level). However, the increase in error is shown in the SI (Fig.~S3) to correlate with the spilling of the associated one-body wavefunction outside the cavity, namely a situation where the PCM approximation becomes more questionable. 

The fully dynamical $GW$@PCM($\omega)$ polarization energies for the HOMO and LUMO energies are reported in Table~\ref{tab:my_label}. The polarization energy per level is rather stable around 1 eV in absolute value. This leads to a gap  closing of $\sim$2 eV from the gas phase to the solvated phase. The errors induced by the $\mathrm{PCM_{PP}}$ scheme are shown to amount to about 1-2 meV, except for the water monomer and hexamer LUMO levels where it increases to $\sim$7-8 meV, in relation with the very large associated spilling.

The accuracy of the $GW$@$\mathrm{PCM_{PP}}$  approximation is very remarkable. This indicates that standard PCM calculations, relying only on the knowledge of the low-frequency reaction field, can be easily extended to include the effect of the dynamics of the solvent electronic degrees of freedom. Furthermore, and as shown in the SI (Fig.~S2), the polarization energy varies rather weakly with the value of the pole energy, with a shift of the order of the meV  for a shift of $\Omega_p$ of the order of the eV. This indicates that a very precise determination of $\Omega_p$ is not a central issue, provided that the low and high frequency limits of $[1 / \epsilon_{\text{opt}}(\omega) - 1]$, or equivalently $\chi^{E}(\omega)$, are  satisfied. 

Along that line, we note that the water pole energy value ($\Omega_p$ = 21 eV)   is close to the classical plasma frequency   $\Omega_{\text{plasmon}}=\sqrt{4\pi n}$ (a.u.)  with $n$ the electron density. For water, $\Omega_{\text{plasmon}}$ amounts indeed to 19.2 eV, accounting for the 8 valence electrons per water molecule, or 21.5 eV including the oxygen 1\textit{s} electrons. The rather large stability of the polarization energy with respect to the plasmon frequency suggests that selecting the classical plasma frequency  may be a simple and accurate strategy in the case of solvents for which no sufficient experimental data for $\epsilon_{\text{opt}}(\omega)$ in the full visible-UV range are available. Such a strategy can also be retrieved by plugging a one-pole model into the f-sum rule. \cite{Hyb86}   

To provide some understanding of the accuracy of the single-plasmon-pole model for $\chi^{E}({\bf r},{\bf r}'; \omega)$ (eq \ref{eqn:xipcm_2}), we note that  in the simple Born model of a unit charge at the center of a spherical cavity of radius $R$, the polarization energy is directly related to $ [ 1 / \epsilon_{\text{opt}} -1]$ through:
$$
P = \frac{e^2}{2R} \left[ \frac{1}{\epsilon_{\text{opt}}}-1 \right]
$$
In the case of a non-spherical cavity, a similar relation holds for the integral of the induced surface charge on the cavity :
$$
\int_{\Gamma} \sigma({\bf x}) \dd{\bf x} = \left[ \frac{1}{\epsilon_{\text{opt}}} -1 \right] Q
$$
with $Q$ the charge in the cavity inducing the distribution $\sigma({\bf x})$ of surface charges in the simplest single-layer IEF approach to PCM.\cite{Cances_1997,Duchemin2016} These two observations shed light on the fact that the effect of the solvent on the monopole of the solute charge redistribution, associated with the photoemission charging process, is directly proportional to $f(\omega,\Omega_p)$. As such,     factoring out the frequency dependence of the response in eq \ref{eqn:xipcm_2} mostly amounts to neglecting the form factor of the cavity and the higher order multipolar (dipole, etc.) contributions from the reaction field. 

\subsection{Accuracy of the PCM$_{\text{COHSEX}}$ \rep{}{adiabatic} scheme}
\begin{figure}[h]
  \includegraphics[width=8.6cm]{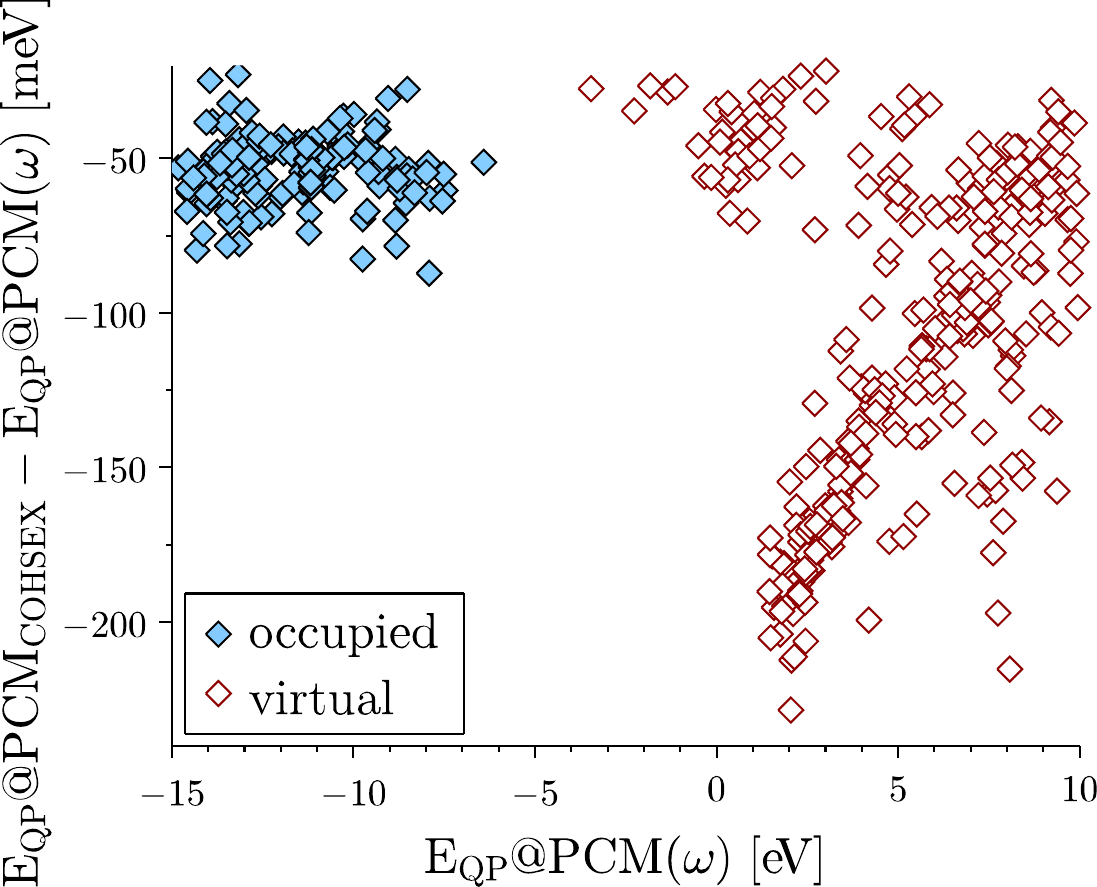} 
  \caption{ Error (meV) on the $GW$@PCM quasiparticle energies associated with the \rep{}{static  (adiabatic  limit)} $\mathrm{PCM_{COHSEX}}$ approach.  The error is taken with respect to the fully dynamical PCM($\omega)$ approach for which the PCM reaction fields are recalculated at each needed frequency. Errors for occupied and virtual energy levels are represented with blue-filled and red-empty diamonds, respectively. }
  \label{fig:cohsex_vs_w}
\end{figure}

Even though including the full dynamics of the solvent electrons can be performed at no cost as compared to  standard PCM calculations in the \rep{BO }{adiabatic} limit, we now explore the accuracy of the $GW$@$\mathrm{PCM_{COHSEX}}$ approach. The reference is again the fully dynamical $GW$@PCM($\omega$) calculation. The error is represented in Fig.~\ref{fig:cohsex_vs_w} for occupied and unoccupied energy levels over a large energy window for the full set of solutes.
The error is further provided in  Table~\ref{tab:my_label} for the HOMO and LUMO levels.

Analyzing Fig.~\ref{fig:cohsex_vs_w}, the error is found to be of the order of {-50 meV} for occupied states and unoccupied states below the vacuum level. With an absolute polarization energy of the order of the electronvolt (in absolute value), this represents a deviation of about 5$\%$.  This is significantly larger than the error associated with the dynamical $\mathrm{PCM_{PP}}$ scheme. Whether such an error should be considered as moderate or large clearly depends on the required accuracy for the problem of interest. For states above the vacuum level, the error can be much larger, with again  a clear correlation of the error  with the spilling outside the cavity (see SI Fig.~S3). For such states, the PCM approximation becomes questionable independently of the treatment of environment electronic degrees of freedom dynamics. 

The systematically negative error indicates that the \rep{BO }{adiabatic} approximation underestimates the polarization energy for the occupied levels but overestimates the correction for the virtual energy levels. In other words, the static approximation leads to ionization potential and electronic affinities in solution that are too large as compared to a fully dynamical PCM calculation. This leads to a partial cancellation of error  when considering the polarization energy associated with the HOMO-LUMO gap, namely the closing in energy of the gap from the gas phase to the solvated phase. 

\begin{figure}[h]
  \includegraphics[width=8.6cm]{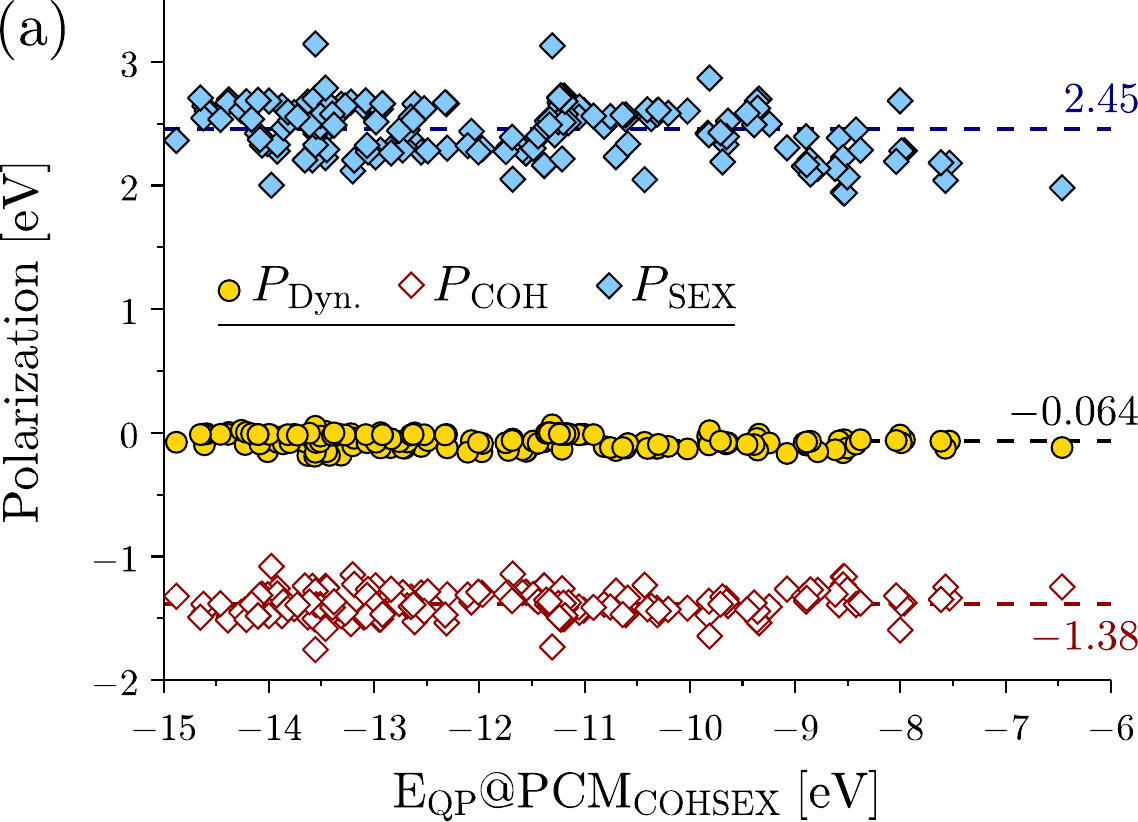} 
  \includegraphics[width=8.6cm]{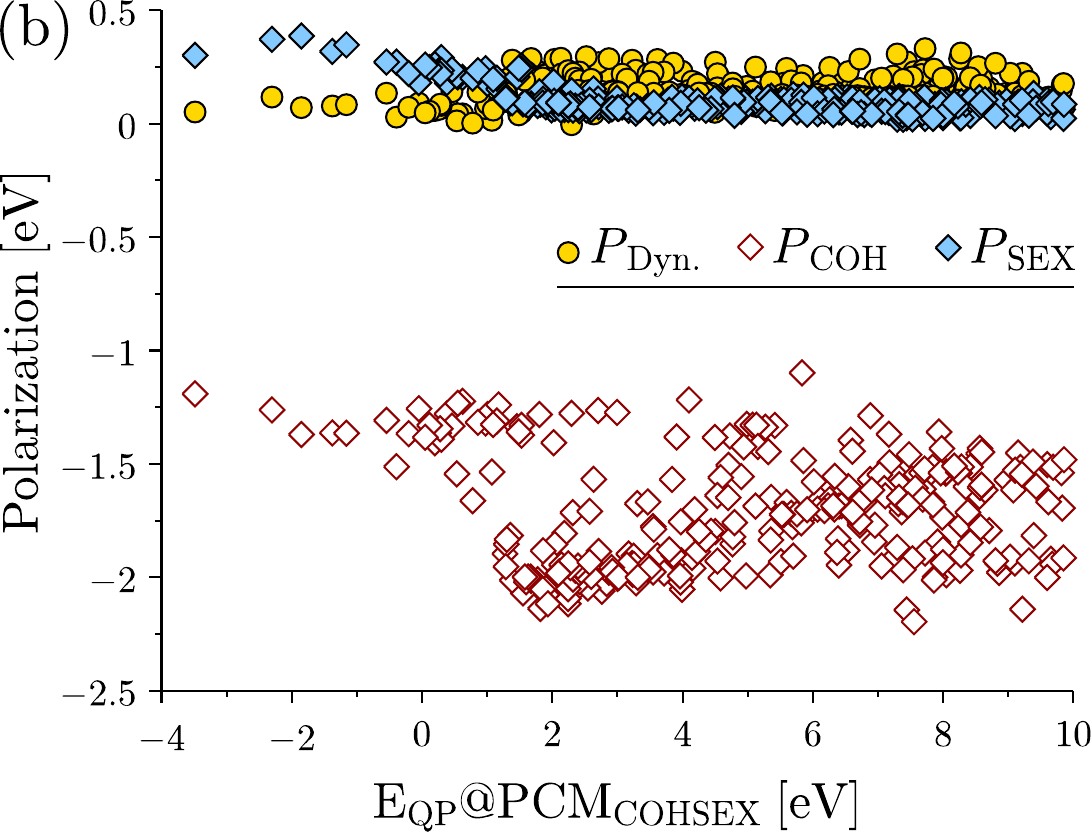}
  \caption{ Decomposition of the polarization energy for (a) occupied and (b) unoccupied energy levels. The $P_{\text{Dyn}}$, $P_{\text{SEX}}$, $P_{\text{COH}}$ correspond to the first, second and third-line contributions from eq \ref{eqn:COHSEX_model}, respectively.  Dashed lines indicate occupied state manifold average values. }
  \label{fig:cohsexdecomp}
\end{figure}

It is interesting to analyze the decomposition reported in eq \ref{eqn:COHSEX_model} partitioning the polarization energy in a SEX, COH and integral renormalization terms. This is represented in Fig.~\ref{fig:cohsexdecomp}. For the occupied states, the renormalization $P_{\text{Dyn}}$ term is seen to be significantly smaller than the perturbative $P_{\text{SEX}}$ and $P_{\text{COH}}$  terms. In the limit of a reaction field very slowly varying over the solute, it was further shown~\cite{Neaton2006} that  the relation $P_{\text{SEX}} = -2 P_{\text{COH}}$ should hold. This relation is only qualitatively verified in the present case. 

Concerning the unoccupied levels, the screened-exchange $P_{\text{SEX}}$ contribution is much smaller, as again justified in the limit of a smooth reaction field. \cite{Neaton2006,Amblard2024} The renormalization $P_{\text{Dyn}}$ term is again significantly smaller than the perturbative contribution $P_{\text{COH}}$ that hardly changes from occupied to unoccupied energy levels. As such, the COH-like term does not contribute much to the renormalization of the HOMO-LUMO gap in solution as compared to the gas phase. Neglecting the COH contribution can thus be qualitatively correct when considering the evolution of energy differences between virtual and unoccupied levels, but fails for the absolute value of the ionization potential and electronic affinities. 

To conclude this exploration of the solvent electron dynamics, we observe that the error associated with the  $\mathrm{PCM_{COHSEX}}$ scheme does not seem to  correlate  with the gap of the molecules (given in the Table S1 of the SI). For sake of illustration, the errors associated, e.g., with the HOMO polarization energies for indigo and benzene-TCNE, the two smallest gap molecules in the test set, are not smaller than the ones associated with the water hexamers that display much larger gaps. For the water solute in a water solvent, one may expect the \rep{BO }{adiabatic} approximation to become questionable since there is no decoupling of energy  between the solute and   solvent electronic degrees of freedom.  However, clearly, the error does not seem to increase significantly. 

Our results are in line with a previous QM/QM' exploration of the embedded $GW$ scheme\cite{Amblard2024}, with an equivalent COHSEX-like treatment of the reaction field, showing that for fullerenes in a fullerene environment, the \rep{}{adiabatic} approximation \rep{that the electrons in the environment react instantaneously to an electronic excitation in the central subsystem}{} would induce errors limited to a few percent as well. Nevertheless, in the case of a metallic environment (a metallic nanotube), the error was shown to increase to a few tenths of an eV for the polarization energy associated with the HOMO and LUMO levels of a water molecule inside the tube. As such, the \rep{}{adiabatic} approximation \rep{of faster solvent electrons}{} seems rather robust, except in extreme situations where the solvent electronic degrees of freedom are characterized by much smaller energies as compared to the solute.  We emphasize however that in any case, a fully dynamical model can be easily set up, with a cost similar to that associated with \rep{a PCM approach }{the standard PCM} where the reaction field is only required in the low-frequency limit \rep{of the electronic dielectric response.}{}

\section{ Conclusions } 

We have studied the impact on the electronic energy level polarization energies of considering explicitly the frequency dependence of the optical dielectric constant of water as a solvent. By polarization energies, we mean the energy levels  shifts from the gas phase to the solvated environment. This study is performed at the $GW$@PCM level, with the $GW$ operator depending explicitly on the  $\epsilon_{\text{opt}}^{-1}({\bf r},{\bf r}';\omega)$ inverse dielectric function through the screened Coulomb potential $W$. Such an explicit frequency dependence offers an ideal playground to explore the impact of the dynamical nature of the solvent optical dielectric response, relying further on a recent parametrization of $\epsilon_{\text{opt}}(\omega)$ in the visible-UV range for water. \cite{Fiedler2020} Accounting for the frequency dependence of the solvent optical dielectric response goes beyond the \rep{}{common adiabatic}  approximation restricting  $\epsilon_{\text{opt}}(\omega)$  to its low frequency $\epsilon_{\infty}$ value. \rep{This restriction is expected to be valid only in the limit where the solvent electrons are assumed to be much faster than that of the solute. Such a limit is labeled the Born-Oppeinheimer (BO) limit in the PCM literature.}{}

In the framework of $GW$@PCM calculations, the \rep{limit of instantaneous solvent electrons response}{ adiabatic PCM approach}   can be properly obtained by assuming that the solvent susceptibility $\chi^{E}(\omega)$ present a pole structure with poles brought to infinity. This yields the so-called static Screened-Exchange and Coulomb-Hole contributions to the solvent reaction field.  This static limit is compared to reference dynamical PCM($\omega)$ calculations where the cavity reaction fields are recalculated explicitly for each needed frequency when building the $GW$ self-energy integral. We observe that the \rep{BO limit }{standard adiabatic approximation} leads to errors of the order of -50 meV for the polarization energy associated with individual energy levels. Due to cancellations of error between occupied and unoccupied levels polarization energies, the error on the HOMO-LUMO gap of the solute is found to be smaller. We emphasize that using directly the low frequency limit of the reaction field, without accounting for the pole structure of the solvent susceptibility, leads to neglecting the Coulomb-hole contribution, with erroneous polarization energies for individual energy levels. 

Besides appraising the error associated with the \rep{instantaneous solvent electronic response}{ static} limit, we have introduced a very simple and accurate approach to dynamical solvents within a single plasmon-pole model for the solvent electronic  susceptibility.
Such a formulation is approximate as it decouples real-space and frequency degrees of freedom but is strictly valid for the monopolar component of the reaction field. As a great advantage, only the low-frequency cavity reaction field \rep{associated with the solvent electronic degrees of freedom}{} needs to be calculated. Finite-frequency reaction fields can further be obtained by simple rescaling. Remarkably, this approach, that we label $GW$@PCM$_{\text{PP}}$ offers an accuracy at the few meV level as compared to the reference PCM($\omega$) calculations. 

The water single-pole frequency ($\Omega_p$=21 eV) was obtained from the best fit of the available experimental $[1/\epsilon_{\text{opt}} -1]$ function. Such a value is close to the classical plasma frequency $\Omega_{\text{plasmon}}=\sqrt{4\pi n}$  (a.u.) associated with water. Further, the $GW$@PCM$_{\text{PP}}$ polarization energies are rather insensitive to the exact $\Omega_p$ value, with a shift of about a meV for an eV change of $\Omega_p$. This suggests that in cases where the full visible/UV $\epsilon_{\text{opt}}(\omega)$ function is not known experimentally for a given solvent, accurate fully dynamical calculations can still be straightforwardly conducted.

Finally, the merging of the $GW$ formalism with the PCM opens the way  to combining the Bethe-Salpeter formalism (BSE) \cite{Strinati1988} with the PCM for the study of optical excitations in systems immersed in a polarizable environment. \cite{Duchemin2018,Kim2024}  The BSE/PCM combination was shown in particular  to account simultaneously for linear-response and state-specific contributions to the solvatochromic shifts. \cite{Duchemin2018} This is an important feature for excitations presenting a hybrid local (Frenkel) and charge-transfer character. Even though the most common BSE implementations rely on the low frequency limit $W(\omega \rightarrow 0)$ of the screened Coulomb potential, dynamical implementations are being explored. \cite{Strinati1982,Strinati1988,Ma2009,Loos2020,Bintrim2022,Yamada2022,Loos2022,Zhang2023,Monino2023}
The present study may facilitate the implementation of a dynamical BSE formalism combined with a dynamical PCM for the environment.

\begin{suppinfo}
 Supporting Information: (I)  a  multiple-pole fit of $\epsilon_{\text{opt}}^{-1}(\omega)$ as an improvement over equation~\ref{eqn:xipcm_1} and in (II) the error for selected polarization energies associated with the single-pole PCM($\omega$) approach with respect to the multiple-fit values, considering   several   single-pole $\Omega_p$ energies. In (III), we evidence the correlation between the errors associated with the $\mathrm{PCM_{PP}}$ and $\mathrm{PCM_{COHSEX}}$ schemes  and the spilling of the orbitals outside the cavity. In (IV), a Table compiling the Kohn-Sham and various ev$GW$@PCM gaps are provided for all molecules of the test set. 
 \end{suppinfo}

\begin{acknowledgement} 
 XB acknowledges numerous discussions with Gabriele D'Avino.
 DA is indebted to ENS Paris-Saclay for his PhD fellowship. 
 This project was provided with computer and storage ressources by GENCI@TGCC thanks to the grants A0130910016 and A0150910016 on the  Joliot-Curie supercomputer (SKL and Rome partitions).
 XB and ID acknowledge support from the French Agence Nationale de la Recherche (ANR) under contract ANR-20-CE29-0005.
\end{acknowledgement} 




\appendix
\section{Appendix: Demonstration of equation \ref{eqn:sigma_c2_1}}
\label{appendix_limits}

In order to tackle the limit $\Omega_p\to\infty$ within eq \ref{eqn:sigma_c2_1}, we first remark that the integral along a closed contour $\mathcal{C}$ enclosing the two poles $\Omega_1$ and $\Omega_2$ of the following product:
\begin{equation}
\int_\mathcal{C} \dd \omega  \cfrac{1}{\omega-\Omega_1}\,\cfrac{1}{\omega-\Omega_2}  =  \cfrac{2i\pi}{\Omega_2-\Omega_1} + \cfrac{2i\pi}{\Omega_1-\Omega_2}=0 
\label{eqn:int_sym}
\end{equation}
is always null.  
We can use this property to rewrite the $GW$ integral by considering only the residues taken at the poles of $G$. Once we have such an expression, it becomes possible to take the limit value of these residues when $\Omega_p\to\infty$. Finally, we can revert the expression by casting the sum over limit values of the residue as an integral of the form $G(E+\omega)\Hp\big[{\widetilde W}(\omega)-{\tilde v}(0)\big]$.

To start our demonstration, we first rewrite the term $[W-{\tilde v}]$ as $\Delta^{(+)}+\Delta^{(-)}$, where $\Delta^{(+)}$ regroup all the poles of $W-{\tilde v}$ within the complex upper-plane while $\Delta^{(-)}$ holds all the  poles of the complex lower-plane. Similarly, we split $G$ into $G_{\text{occ}}$ and $G_{\text{vir}}$ with poles occupying also respectively the upper-plane and the lower-plane. 
\begin{equation}
\begin{split}
   \cfrac{i}{2\pi} \int_{-\infty}^\infty \dd \omega \; e^{i\omega\eta}\, & G(E+\omega)\Hp\big[W(\omega)-{\tilde v}(\omega)\big] \\
=  \cfrac{i}{2\pi} \int_{-\infty}^\infty \dd \omega \; e^{i\omega\eta}\, \Big( & \, G_{\text{occ}}(E+\omega) + G_{\text{vir}}(E+\omega) \, \Big) \\
\times \; \Big( & \, \Delta^{(+)}(\omega) + \Delta^{(-)}(\omega) \, \Big).
\end{split}
\label{eqn:app_2}
\end{equation}
We insist on the fact that all these sub-quantities have a $1/\omega$ asymptotic behavior, with no constant terms, as they are only composed of simple poles. Thus, expression \ref{eqn:app_2} is clearly an integral over products of simple poles. Using eq \ref{eqn:int_sym} as selection rule for which of these product residues to keep after integration, we can now express the equation \ref{eqn:sigma_c2_1} as:
\begin{equation}
\begin{split}
  \cfrac{i}{2\pi} \int_{-\infty}^\infty \dd & \omega \; e^{i\omega\eta}\, G(E+\omega)\Hp\big[W(\omega)-{\tilde v}(\omega)\big] \\
= & -\!\! \sum_i \mathrm{Res}(G_{\text{occ}},\varepsilon_i) \, \Delta^{(-)}(\varepsilon_i-E) \\
  & - \!\!\sum_k G_{\text{vir}}(E+\Omega_k) \, \mathrm{Res}(\Delta^{(+)},\Omega_k).
\end{split}
\label{eqn:app_2_2}
\end{equation}
$\mathrm{Res}(f,x)$ corresponds to the residue of the function $f$, associated to its pole $x$. We can then use eq \ref{eqn:int_sym} once again to rewrite the sum over $k$ as a sum over the empty states $a$ as:
\begin{equation}
\begin{split}
  &   \!\!\sum_k G_{\text{vir}}(E+\Omega_k) \, \mathrm{Res}(\Delta^{(+)},\Omega_k) \\
= - &  \!\! \sum_a \mathrm{Res}(G_{\text{vir}},\varepsilon_a) \, \Delta^{(+)}(\varepsilon_a-E), \\
\end{split}
\end{equation}
leading to an expression of eq \ref{eqn:sigma_c2_1} that involves only residues taken at the poles of $G$:
\begin{equation}
\begin{split}
   \cfrac{i}{2\pi} \int_{-\infty}^\infty \dd & \omega \; e^{i\omega\eta}\,G(E+\omega)\Hp\big[W(\omega)-{\tilde v}(\omega)\big] \\
= & -\!\! \sum_i \mathrm{Res}(G_{\text{occ}},\varepsilon_i) \, \Delta^{(-)}(\varepsilon_i-E) \\
  & +\!\! \sum_a \mathrm{Res}(G_{\text{vir}},\varepsilon_a) \, \Delta^{(+)}(\varepsilon_a-E).
\end{split}
\label{eqn:app_4}
\end{equation}
Now that we have an expression that requires only evaluation of $\Delta^{(+)}$ and $\Delta^{(-)}$ at finite value of $\omega$, we can explore their limit when $\Omega_p\to\infty$. In order to do so, we remark that for any finite value $\omega$, we can write:
\begin{equation}
\begin{split}
\lim_{\Omega_p\to\infty} W(\omega) - {\tilde v}(\omega)   
= & \lim_{\Omega_p\to\infty} {\tilde v}(\omega) \cdot {\tilde \chi}(\omega) \cdot {\tilde v}(\omega) \\
 = & \;  {\tilde v}(0) \cdot {\tilde \chi}(\omega) \cdot {\tilde v}(0), \\
\end{split}
\end{equation}
with ${\tilde \chi}(\omega)$ also taken at the limit:
\begin{equation}
    {\tilde \chi}(\omega) = \chi_0^\sys(\omega)+ \chi_0^\sys(\omega)\cdot{\tilde v }(0)\cdot{\tilde \chi}(\omega). \label{eqn:tildexi_stat}
\end{equation}
Thus, we can write 
\begin{equation}
\begin{split}
\lim_{\Omega_p\to\infty} W(\omega) - {\tilde v}(\omega)   
= {\widetilde W}(\omega) - {\tilde v}(0), 
\end{split}
\label{eqn:app_7}
\end{equation}
with 
\begin{equation}
\begin{split}
{\widetilde W}(\omega) = {\tilde v}(0) + {\tilde v}(0) \cdot \chi_0^\sys(\omega) \cdot {\widetilde W}(\omega). \\
\end{split}
\label{eqn:app_8}
\end{equation}
The screened Coulomb integrals ${\widetilde W}$ correspond thus to the screened PCM Coulomb integrals obtained within the static limit of the solvent response, and renormalized by the dynamical solute free susceptibility. We can then proceed as previously by separating ${\widetilde W}(\omega) - {\tilde v}(0)={\tilde \Delta}^{(+)}(\omega) +{\tilde \Delta}^{(-)}(\omega)$ into contributions of the upper and lower-plane poles. The equation \ref{eqn:app_7} being true for all finite $\omega$, this implies that we now can make the identification:  
\begin{equation}
\begin{split}
\lim_{\Omega_p\to\infty} \Delta^{(+/-)}(\omega) = {\tilde \Delta}^{(+/-)}(\omega).
\end{split}
\end{equation}
The limit values of $\Delta^{(+/-)}(\omega)$ can be re-injected into eq \ref{eqn:app_4}: 
\begin{equation}
\begin{split}
  \lim_{\Omega_p\to\infty} \cfrac{i}{2\pi} \int_{-\infty}^\infty & \dd \omega \; e^{i\omega\eta}\,G(E+\omega)\Hp\big[W(\omega)-{\tilde v}(\omega)\big] \\
= & -\!\! \sum_i \mathrm{Res}(G_{\text{occ}},\varepsilon_i) \, {\tilde \Delta}^{(-)}(\varepsilon_i-E) \\ 
  & +\!\! \sum_a \mathrm{Res}(G_{\text{vir}},\varepsilon_a) \, {\tilde \Delta}^{(+)}(\varepsilon_a-E). \\
\end{split}
\label{eqn:app_11}
\end{equation}
It is now possible to revert the derivation starting from eq \ref{eqn:app_11} and rewinding the steps \ref{eqn:app_4} back to \ref{eqn:app_2} to finally obtain that:
\begin{equation}
\begin{split}
  \lim_{\Omega_p\to\infty}\;\; & \cfrac{i}{2\pi} \int_{-\infty}^\infty \dd \omega \; e^{i\omega\eta}\,G(E+\omega)\Hp\big[W(\omega)-{\tilde v}(\omega)\big] \\
 = \; & \cfrac{i}{2\pi} \int_{-\infty}^\infty \dd \omega \; e^{i\omega\eta}\,G(E+\omega)\Hp\big[{\widetilde W}(\omega)-{\tilde v}(0)\big]. \\
\end{split}
\end{equation}

\bibliography{xavbib.bib}

\begin{mcitethebibliography}{104}
\bibitem[pcm()]{pcmnotations}
It may seem somehow awkward to use the notation $\epsilon_{\infty}$ for the low frequency limit of the macroscopic dielectric constant in the optical range. However, in the PCM language, the frequencies associated with the electronic degrees of freedom remain always much larger than that associated with the slow degrees of freedom contributing to $\epsilon_0$.\relax
\mciteBstWouldAddEndPunctfalse
\mciteSetBstMidEndSepPunct{\mcitedefaultmidpunct}
{}{\mcitedefaultseppunct}\relax
\EndOfBibitem
\end{mcitethebibliography}

\end{document}